\documentclass[aps,prd,twocolumn,floatfix,showpacs,superscriptaddress,showkeys]{revtex4-1}
\usepackage{amsmath,amsfonts,amssymb,bm}
\usepackage{graphicx}
\usepackage{hyperref}
\usepackage{color}
\usepackage{soul}
\usepackage{subfigure}
\usepackage{multirow}
\usepackage{textcomp}
\usepackage[USenglish]{babel}
\usepackage{slashed}
\usepackage[utf8]{inputenc}
\usepackage[bottom]{footmisc}

\begin{document}

\title{Fermion propagator in a rotating environment}
\author{Alejandro Ayala}
\affiliation{Instituto de Ciencias Nucleares, Universidad Nacional Aut\'onoma de M\'exico, Apartado Postal 70-543, CdMx 04510, Mexico.}
\affiliation{Centre for Theoretical and Mathematical Physics, and Department of Physics, University of Cape Town, Rondebosch 7700, South Africa.}

\author{L. A. Hernández}
\affiliation{Centre for Theoretical and Mathematical Physics, and Department of Physics, University of Cape Town, Rondebosch 7700, South Africa.}
\affiliation{Departamento de F\'isica, Universidad Aut\'onoma Metropolitana-Iztapalapa, Av. San Rafael Atlixco 186, C.P, CdMx 09340, Mexico.}
\affiliation{Facultad de Ciencias de la Educaci\'on, Universidad Aut\'onoma de Tlaxcala, Tlaxcala, 90000, Mexico.
}

\author{K. Raya}
\affiliation{Instituto de Ciencias Nucleares, Universidad Nacional Aut\'onoma de M\'exico, Apartado Postal 70-543, CdMx 04510, Mexico.}
\affiliation{School  of  Physics,  Nankai  University,  Tianjin  300071,  China.}

\author{R. Zamora}
\affiliation{Instituto de Ciencias B\'asicas, Universidad Diego Portales, Casilla 298-V, Santiago, Chile.}
\affiliation{Centro de Investigaci\'on y Desarrollo en Ciencias Aeroespaciales (CIDCA), Fuerza A\'erea de Chile, Casilla 8020744, Santiago, Chile.}


\begin{abstract}
We apply the exponential operator method to derive the propagator for a fermion immersed within a rigidly rotating environment with cylindrical geometry. Given that the rotation axis provides a preferred direction, Lorentz symmetry is lost and the general solution is not translationally invariant in the radial coordinate. However, under the approximation that the fermion is completely dragged by the vortical motion, valid for large angular velocities, translation invariance is recovered. The propagator can then be written in momentum space. The result is suited to be used applying ordinary Feynman rules for perturbative calculations in momentum space.
\end{abstract}

\keywords{Fermion propagator, rotation, polarization}
\maketitle


\section{Introduction}\label{intro}

Collisions of heavy nuclei at high energies produce deconfined strongly interacting matter, dubbed as the quark-gluon plasma (QGP). When these collisions are off-center, the inhomogeneity of the matter distribution in the transverse plane causes the colliding region to develop an orbital angular velocity $\Omega$ directed along the normal to the reaction plane~\cite{Becattini:2007sr,Becattini:2015ska}. Estimates of this angular velocity provide a value $\Omega\sim 10^{22}$ s$^{-1}$~\cite{STAR:2017ckg}. 

When the vortical motion is transferred to the particles spin within the QGP, its effect can show, upon hadronization, as a global hadron polarization, namely, a preferred direction of the spin of hadrons along the normal to the reaction plane. Recent measurements of the global $\Lambda$ and $\overline{\Lambda}$ polarizations as functions of collision energy~\cite{STAR:2017ckg,Abelev:2007zk,Adam:2018ivw}
show that the $\overline{\Lambda}$ polarization rises more steeply than the $\Lambda$ polarization when the collision energy decreases. A suitable explanation of this intriguing result motivates the search for the conditions to align the particle's spin to the global vortical motion. In particular, it is important to establish how these conditions depend on parameters such as the collision energy $\sqrt{s_{NN}}$, the impact parameter $b$, the temperature $T$, the baryon chemical potential $\mu_B$ and the global angular velocity $\Omega$.

The problem has attracted a great deal of attention over the last several years~\cite{Becattini:2013vja,Xie:2015xpa,Sorin:2016smp,Sun:2017xhx,Suvarieva:2017pmf,Xie:2016fjj,Li:2017slc,Xia:2017iav,Karpenko:2018erl,Suvarieva:2018wmh,Han:2017hdi,Xia:2018tes,Baznat:2017jfj,Kolomeitsev:2018svb,Xie:2019jun,Guo:2019joy,Ma:2019mrv,Kapusta:2020npk}.
In a recent work~\cite{Ayala:2020soy}, we have explored the $\Lambda$ and $\overline{\Lambda}$ polarization within a model where the overlap region in a peripheral heavy-ion collision consists of a dense core and a less dense corona, from where different $\Lambda$ and $\overline{\Lambda}$ production mechanisms are at play~\cite{Ayala:2001jp}. The calculation relies on the computation of the relaxation time that a strange quark or antiquark takes to align its spin to the global vorticity at finite temperature and baryon density, which was computed introducing a phenomenological coupling between the quark spin and the thermal vorticity~\cite{Ayala:2019iin,Ayala:2020ndx}.

However, in order to obtain a better estimate, it is important to set up the problem in terms of a first principles calculation to see whether the intuitive use of the above mentioned phenomenological coupling is correct. In this work we take the first step toward achieving this goal and compute the propagator of fermions within a rotating environment. 

Fermion and scalar propagators within a rotating system and subject to a thermal bath have been first computed in Ref.~\cite{Vilenkin:1980zv}. In the absence of medium effects,  Lorentz symmetry is still broken due to the preferred direction provided by the angular velocity. Therefore the propagators are in general given by cumbersome expressions in the space-time representation. The situation worsens even further when including finite temperature as well as chemical potential effects, since then analytical results are not possible and the calculation requires a numerical estimate. This approach has been taken in Ref.~\cite{Wei:2020xfd} to compute the rotation effects on meson masses.  It is thus desirable to find an expression that, under suitable conditions, can be approximated by a translationally invariant result. As we show in this work, this can be done provided we keep the first nontrivial contribution in the angular velocity $\Omega$, which is taken as a large quantity compared to the expansion rate $\Gamma$, effectively making fermions partake of the rigid rotational motion. In this approximation the focus is on how the rotation influences the spin states rather than on the detailed dynamics of the fermion motion. In this work we concentrate on the computation of the fermion propagator in vacuum, and postpone the discussion of medium effects for a future work. 

To compute the fermion propagator subject to rotation, we follow the method introduced in Refs.~\cite{Iablokov:2019rpd,Iablokov:2020upc} that requires knowledge of the explicit set of solutions of the Dirac equation. These solutions have been studied by several authors imposing different boundary conditions. Working in a cylindrical geometry, in the pioneering work that introduced the MIT bag model~\cite{Chodos:1974je}, these boundary conditions are chosen such that the fermion current normal to the cylinder surface vanishes. These conditions are nowadays known as the MIT boundary conditions. A slight modification of these conditions, known as the chiral MIT conditions, can also be imposed on the fermion modes~\cite{Chernodub:2016kxh}. Bound and unbound solutions have also been studied in Ref.~\cite{Ambrus:2015lfr}. The solutions can also be found in the presence of a magnetic field pointing in the same direction as the angular velocity, given that the geometry of the problem is not modified by the presence of the field~\cite{Chernodub:2016kxh,Chen:2015hfc,Ebihara:2016fwa}. Thermal and rotating states
were studied in Ref.~\cite{Ambrus:2014uqa}. Lattice QCD has also been formulated in rotating frames to study the angular momenta of gluons and quarks in a rotating QCD vacuum~\cite{Yamamoto:2013zwa}. In all these calculations, the causality condition, whereby the angular velocity and the cylinder radius $R$ must satisfy $R\Omega<1$, is imposed.

In this work we take a pragmatic approach. We find the solutions to the Dirac equation for fermions rigidly rotating inside a cylinder. In order to satisfy the causality condition for a given $\Omega$, the solutions are taken as not existent for $r>R$, but otherwise do not need to satisfy a given boundary condition. The problem thus formulated, lends itself to attempt finding the fermion propagator in momentum space, which is a useful quantity to employ in perturbative calculations using ordinary Feynman rules. The work is organized as follows: In Sec.~\ref{SecII} we formulate the Dirac equation and find the solutions for a rigidly rotating cylinder in unbound space. In Sec.~\ref{SecIII} we find the fermion propagator. We apply the approximation whereby fermions are totally dragged by the rigid motion to find the expression of this propagator in momentum space. We finally summarize and provide an outlook for the use of these results in Sec.~\ref{Concl}.

\section{Fermions  in a rigidly rotating cylinder}\label{SecII}

The physics within a relativistic rotating frame is most easily described in terms of a metric tensor resembling that of a curved space-time. For our purposes, we consider that the interaction region after a relativistic heavy-ion collision can be thought of as a rigid 
cylinder rotating around the $\hat{z}$-axis with constant angular velocity $\Omega$. Therefore, the metric tensor is given by
\begin{equation}\label{eq:metric}
    g_{\mu\nu} = \begin{pmatrix}
1-(x^2+y^2)\Omega^2 & y \Omega & -x \Omega & 0 \\
y \Omega & -1 & 0 & 0 \\
-x \Omega & 0 & -1 & 0 \\
0 & 0 & 0 & -1\end{pmatrix}.
\end{equation}
A fermion with mass $m$ within the cylinder is described by the Dirac equation \cite{Chen:2015hfc,Chernodub:2016kxh}
\begin{equation}\label{eq:DiracEc1}
    \left[i(\gamma^\mu \partial_\mu +\Gamma_\mu)-m\right]\psi = 0,
\end{equation}
where $\Gamma_\mu$ corresponds to the affine connection, determined from the equations
\begin{eqnarray}\label{eq:AffineC}
\Gamma_\mu &=& -\frac{i}{4}\omega_{\mu i j} \sigma^{ij},\nonumber\\
\omega_{\mu i j}&=&g_{\alpha \beta} e_i^\alpha(\partial_\mu e_j^\beta + \Gamma_{\mu \nu}^\beta e_j^\nu),
\end{eqnarray}
where the commutator
\begin{eqnarray}
\sigma^{ij}=\frac{i}{2}[\gamma^i, \gamma^j]
\end{eqnarray}
corresponds to the fermion spin
and, the Christoffel symbols are given in terms of the metric tensor by
\begin{equation}\label{eq:ChristoffelC}
\Gamma_{\mu\nu}^\lambda=\frac{1}{2}g^{\lambda \sigma}(g_{\sigma \nu,\mu}+g_{\mu \sigma,\nu}-g_{\mu \nu,\sigma}).
\end{equation}
Greek indices ($\mu,\nu,\ldots=t,x,y,z$) refer to the general coordinates in the moving frame, while Latin indices ($i,j,\ldots=0,1,2,3$) refer to the Cartesian coordinates in the local rest frame. Notice that $\gamma^\mu = e_i^\mu \gamma^i$ corresponds to the Dirac matrices in curved space-time, which satisfy the usual anticommutation relations
\begin{equation}
    \{ \gamma^\mu, \gamma^\nu \} = 0.
\end{equation}
The tetrad $e_i^\mu$ is written in the Cartesian gauge~\cite{Ambrus:2014uqa}, such that it connects the general coordinates with the Cartesian coordinates in the local rest frame as $x^\mu = e_i^\mu x^i$. Explicitly,
\begin{eqnarray}
e^t_0 &=& e^x_1 = e^y_2 = e^z_3 = 1,\nonumber\\
e^t_1 &=& y\Omega,\nonumber\\e^t_2&=&-x\Omega,
\end{eqnarray}
with the rest of the components being equal to zero. The nonzero components of the Christoffel symbols, Eq.~\eqref{eq:ChristoffelC}, are
\begin{eqnarray}\nonumber
\Gamma_{tx}^y &=& \Gamma_{xt}^y=\Omega\;,\;\Gamma_{ty}^x = \Gamma_{yt}^x=-\Omega\;,\\
\Gamma_{tt}^x &=& -x \Omega^2\;,\;\Gamma_{tt}^y=-y\Omega^2\;.
\end{eqnarray}
Thus, given the above results and Eq.~\eqref{eq:AffineC}, it is straightforward to see that $\Gamma_\mu$ merely reduces to
\begin{equation}\label{eq:AffineC2}
\Gamma_\mu \to \Gamma_t = -\frac{i}{2}\sigma^{12}\;.
\end{equation}
Subsequently, the gamma matrices in the rotating frame are expressed, in terms of the usual gamma matrices, as
\begin{eqnarray}\nonumber
\gamma^t = \gamma^0\;,\;\gamma^x = \gamma^1 + y \Omega \gamma^0\;,\\
\gamma^z = \gamma^3\;,\;\gamma^y = \gamma^2 - x \Omega \gamma^0\;\;.
\end{eqnarray}
Therefore, Eq.~\eqref{eq:DiracEc1} becomes
\begin{eqnarray}\nonumber
\Big[i\gamma^0\left(\partial_t - x \Omega \partial_y + y\Omega \partial_x-\frac{i}{2}\Omega \sigma^{12} \right) &\\
 +i\gamma^1\partial_x + i \gamma^2 \partial_y + i \gamma^3 \partial_z -m\Big]\psi &= 0\;.\label{eq:DiracEc2}
\end{eqnarray}
In the Dirac representation, 
\begin{equation}\label{eq:sigma12}
    \sigma^{12} = \begin{pmatrix}
    \sigma_3 & 0 \\
    0 & \sigma_3
    \end{pmatrix},
\end{equation}
where $\sigma_3 = \text{diag}(1,\;-1)$ is the Pauli matrix associated with the third component of the spin. In consequence, Eq.~\eqref{eq:DiracEc2} can be conveniently rewritten as
\begin{eqnarray}\label{DiracEc3}
   \left[\gamma^0(i\partial_t+\Omega \hat{J}_z)+i\vec{\gamma}\cdot \vec{\nabla}-m \right]\psi=0,
\end{eqnarray}
where
\begin{eqnarray}
\label{eq:Jdef1}
   \hat{J}_z\equiv \hat{L}_z+\hat{S}_z= -i(x  \partial_y - y \partial_x)+\frac{1}{2} \sigma^{12}.
\end{eqnarray}
$\hat{J}_z$ defines $\hat{z}$-component of the total angular momentum operator, such that the first term is associated with the orbital angular momentum ($\hat{L}_z$), while the second one is related with the spin ($\hat{S}_z$). As usual, $-i\vec{\nabla}$ is the momentum operator. The solution of Eq.~\eqref{DiracEc3} has been studied in many works, e.g.~\cite{Ambrus:2014uqa,Ambrus:2015lfr,Chen:2015hfc,Jiang:2016wvv,Chernodub:2016kxh}. For instance, Ref.~\cite{Ambrus:2015lfr} thoroughly discusses the bound and unbound solutions. 

At this stage, one could in principle be tempted to write Eq.~\eqref{DiracEc3} already in cylindrical coordinates. Nevertheless, in such cases, the relevant Dirac matrices become coordinate dependent and use of some not uniquely determined unitary transformations would be required~\cite{Loewe:2011gs,DirRMF}. To avoid this issue, we follow a different strategy. Consider a solution of the form
\begin{equation}\label{eq:psiphi}
    \psi(x) =  \left[\gamma^0(i\partial_t+\Omega \hat{J}_z)+i\vec{\gamma}\cdot \vec{\nabla}+m \right]\phi(x)\;.
\end{equation}
Then, Eq.~\eqref{DiracEc3} implies that $\phi(x)$ obeys the second order differential equation
\begin{equation}
\label{eq:forphi1}
    \left[(i\partial_t+\Omega \hat{J}_z)^2 + \partial_x^2 + \partial_y^2+\partial_z^2 -m^2 \right]\phi(x)=0\;.
\end{equation}
Since the above equation does not contain gamma matrices, to find solutions consistent with the background geometry it now becomes convenient to work in cylindrical coordinates, $(t,x,t,z) \to (t,\rho \sin\varphi,\rho \cos\varphi,z)$. Thus, Eq.~\eqref{eq:forphi1} becomes
\begin{equation}
\label{eq:forphi2}
    \!\!\left[(i\partial_t+\Omega \hat{J}_z)^2 \!+\! \left(\partial_\rho^2+ \frac{1}{\rho}\partial_\rho + \frac{1}{\rho^2}\partial_\varphi^2\right) \!+\partial_z^2-m^2 \right]\!\phi(x)\!=\!0.
\end{equation}
In these coordinates 
\begin{equation}
 \hat{L}_z=-i\partial_\varphi \Rightarrow \hat{J}_z = -i \partial_\varphi + \hat{S}_z\;.   
\end{equation}
Assuming that the solution of Eq.~\eqref{eq:forphi2} admits a separation of variables, we can write
\begin{equation}
    \phi(x)=e^{-iEt+i k_z z }u(\rho,\varphi).
\end{equation}
Due to the form of $\hat{S}_z$, Eqs.~\eqref{eq:sigma12} and \eqref{eq:Jdef1}, the spin operator will produce eigenvalues $s=\pm 1/2$.  Consequently, total angular momentum conservation ($j=\ell+s$) demands solutions with $\ell$ (for $s=+1/2$) and $\ell+1$ (for $s=-1/2$). Thus, writing 
\begin{eqnarray}
\phi=\left(
\begin{array}{c}
\phi_1\\ \phi_2\\ \phi_3\\ \phi_4
\end{array}\right),
\label{vecphi}
\end{eqnarray}
Eq.~(\ref{eq:forphi2}) becomes
\begin{eqnarray}\nonumber
&\Big[\left(E+(l+\frac{1}{2})\Omega\right)^2+ \left(\partial_\rho^2+ \frac{1}{\rho}\partial_\rho - \frac{\ell^2}{\rho^2}\right)\\
&
-k_z^2-m^2 \Big] \phi_{1,3}(x)=0,\\
&\Big[\left(E+(l+1-\frac{1}{2})\Omega\right)^2+ \left(\partial_\rho^2+\nonumber \frac{1}{\rho}\partial_\rho - \frac{(\ell+1)^2}{\rho^2}\right)\\
&-k_z^2-m^2 \Big] \phi_{2,4}(x)=0.
\end{eqnarray}
The above correspond to Bessel equations
\begin{eqnarray}
\label{eq:Bessel1}
\Big[\rho^2 \partial_\rho^2 + \rho \partial_\rho +(\rho^2  k_\perp^2-\ell^2)\Big]\phi_{1,3}&=0\;, \\
\label{eq:Bessel2}
\Big[\rho^2 \partial_\rho^2 + \rho \partial_\rho +(\rho^2  k_\perp^2-(\ell+1)^2)\Big]\phi_{2,4}&=0\;,
\end{eqnarray}
where
\begin{eqnarray}
k_\perp^2=\tilde{E}^2-k_z^2-m^2,
\label{onmassshell}
\end{eqnarray}
is the transverse momentum squared and we have defined $\tilde{E}\equiv E+j\: \Omega$, which represents the fermion energy as seen from the inertial frame. The solutions of Eqs.~\eqref{eq:Bessel1} and~\eqref{eq:Bessel2} that are finite for $\rho \to 0$ are given by Bessel functions of the first kind, which means that 
\begin{eqnarray}
u(\rho,\varphi)&=&e^{i\varphi \ell} J_\ell(k_\perp \rho)\;,\;\text{for $\phi_{1,3}$}\;,\\
u(\rho,\varphi)&=&e^{i\varphi (\ell+1)} J_{\ell+1}(k_\perp \rho)\;,\;\text{for $\phi_{2,4}$}\;.
\end{eqnarray}
Therefore, the solution of Eq.~(\ref{eq:psiphi}) can be explicitly written as
\begin{eqnarray}
\phi(x)=\begin{pmatrix}
J_\ell(k_\perp \rho) \\
J_{\ell+1}(k_\perp \rho)e^{i\varphi} \\
J_\ell(k_\perp \rho) \\
J_{\ell+1}(k_\perp \rho)e^{i\varphi}\end{pmatrix} e^{-iEt+ik_z z +i\ell \varphi }.
\label{explphi}
\end{eqnarray}

Having determined the solutions $\phi$, Eq.~\eqref{eq:psiphi} can be used to find the spinor wave functions which become
\begin{eqnarray}\nonumber
\psi(x) &=& \begin{pmatrix}
\tilde{E}+m & 0 & -k_z & -\mathcal{P}_{-} \\
0 & \tilde{E}+m & -\mathcal{P}_{+} & k_z \\
k_z & \mathcal{P}_{-} & -\tilde{E}+m & 0 \\
\mathcal{P}_{+} & -k_z & 0 & -\tilde{E}+m\end{pmatrix} \\ \label{eq:psiphi2}
&\times& 
\begin{pmatrix}
J_\ell(k_\perp \rho) \\
J_{\ell+1}(k_\perp \rho)e^{i\varphi} \\
J_\ell(k_\perp \rho) \\
J_{\ell+1}(k_\perp \rho)e^{i\varphi}\end{pmatrix} e^{-iEt+ik_z z +i\ell \varphi }\;,
\end{eqnarray}
where $\mathcal{P}_{\pm}=k_x\pm i k_y$. In cylindrical coordinates,
\begin{equation}\label{eq:Ppm}
    \mathcal{P}_{\pm}=-i e^{\pm i\varphi}(\partial_\rho \pm i\rho^{-1} \partial_\varphi)\;.
\end{equation}
$\mathcal{P}_{\pm}$ acts on the wave functions as ladder operators~\cite{Ambrus:2015lfr}, namely
\begin{equation}\label{eq:Ppm2}
    \mathcal{P}_{\pm} e^{i\ell \varphi}J_{\ell}(k_\perp \rho)= \pm i k_\perp e^{i(\ell \pm 1)\varphi}J_{\ell \pm 1}(k_\perp \rho)\;.
\end{equation}
Thus, combining Eqs.~\eqref{eq:psiphi2}-\eqref{eq:Ppm2}, the explicit  result for $\psi(x)$ reads as
\begin{eqnarray}\nonumber
    \psi(x) &=& \begin{pmatrix}
    [\tilde{E}+m-k_z + i k_\perp]J_\ell(k_\perp \rho) \\
    [\tilde{E}+m+k_z - i k_\perp]J_{\ell+1}(k_\perp \rho)e^{i\varphi} \\
   [-\tilde{E}+m+k_z - i k_\perp]J_\ell(k_\perp \rho) \\
    [-\tilde{E}+m-k_z + i k_\perp]J_{\ell+1}(k_\perp \rho)e^{i\varphi} \\
    \end{pmatrix}\\
    &\times&e^{-i\tilde{E}t+ik_z z +i\ell \varphi }.\label{eq:SolPhi}
\end{eqnarray}
Armed with the explicit expressions, we now follow the approach discussed in Refs.~\cite{Iablokov:2020upc,Iablokov:2019rpd} to find the fermion propagator.

\section{Fermion propagator in a rigidly rotating cylinder}\label{SecIII}

Recall that in order to find the solution for an equation describing the Green's function $G(x,x')$ of a given differential operator $H(\partial_x,x)$, namely
\begin{eqnarray}
H(\partial_x,x)G(x,x')=\delta^4(x-x'),
\label{propeq}
\end{eqnarray}
the Green's function can be represented as
\begin{eqnarray}
G(x,x')=(-i)\int_{-\infty}^0 d\tau\ U(x,x';\tau),
\label{represented}
\end{eqnarray}
where $\tau$ is known as a proper time parameter and $U(x,x';\tau)$ is an evolution operator in this proper-time. This operator satisfies
\begin{eqnarray}
i\partial_\tau U(x,x';\tau)=H(\partial_x,x)U(x,x';\tau),
\label{evolutionop}
\end{eqnarray}
together with the boundary conditions
\begin{eqnarray}
U(x,x';-\infty)&=&0,\nonumber\\
U(x,x';0)&=&\delta^4(x-x'),
\label{boundcond}
\end{eqnarray}
from where the  solution is readily found as
\begin{eqnarray}
U(x,x';\tau)=\exp[-i\tau H(\partial_x,x)]\delta^4(x-x').
\label{explU}
\end{eqnarray}

In order to find the precise form of the proper-time evolution operator, we can use that, when the eigenfunctions $\phi_\lambda(x)$ of the operator $H(\partial_x,x)$ are known, the Dirac delta-function can be expressed in terms of the closure relation obeyed by the eigenfunctions $\phi_\lambda(x)$, namely
\begin{eqnarray}
\sum_\lambda\phi_\lambda(x)\phi^\dagger_\lambda(x')=\delta^4(x-x').
\label{closure}
\end{eqnarray}
Therefore, an exact expression for the proper-time evolution operator can be written as
\begin{eqnarray}
U(x,x';\tau)=\sum_\lambda\exp[-i\tau\lambda]\phi_\lambda(x)\phi^\dagger_\lambda(x'),
\label{evolexpl}
\end{eqnarray}
where we have used the eigenvalue equation
\begin{eqnarray}
H(\partial_x,x)\phi_\lambda(x)=\lambda\phi_\lambda(x).
\label{eigeneq}
\end{eqnarray}
Using Eqs.~(\ref{represented}) and~(\ref{evolexpl}), the propagator $G(x,x')$ can be written as
\begin{eqnarray}
\!\!\!\!\!\!\!G(x,x')=(-i) \int_{-\infty}^0d\tau\sum_\lambda\exp[-i\tau\lambda]\phi_\lambda(x)\phi^\dagger_\lambda(x').
\label{propG}
\end{eqnarray}

It is easy to show that the solutions in Eq.~(\ref{explphi}) satisfy the closure relation
\begin{eqnarray}
\sum_{l=-\infty}^{\infty} \int \frac{dE dk_z dk_\perp k_\perp}{(2\pi)^3}\phi(x)\phi^\dagger(x')=\delta^4(x-x'),
\label{closureforus}
\end{eqnarray}
where we have taken the {\it quantum numbers} $E,k_\perp,k_z,\ell$ as independent, namely, the on-shell restriction of Eq.~(\ref{onmassshell}) is not imposed, as corresponds for a procedure to find the propagator. Furthermore, notice that $k_\perp$ is taken in the continuous domain $0\leq k_\perp \leq\infty$ and thus, no boundary restriction is required on the space variable $\rho$.

To obtain the fermion propagator, we notice that, in the same manner that the solutions of the Dirac equation are obtained from the solutions to the second order differential equation, Eq.~(\ref{eq:psiphi}), the fermion propagator $S(x,x')$ can be  derived~\cite{Iablokov:2019rpd} from
\begin{eqnarray}
\!\!\!\!\!\!S(x,x')=\left[\gamma^0(i\partial_t+\Omega \hat{J}_z)+i\vec{\gamma}\cdot \vec{\nabla}+m \right]G(x,x'),
\label{ferpropfromscalar}
\end{eqnarray}
where
\begin{eqnarray}
G(x,x')&=&(-i) \int_{-\infty}^0 d\tau e^{-i \tau (\tilde{E}^2-k_\perp^2-k_z^2-m^2+i\epsilon)}\nonumber\\ &\times&\sum_{\ell=-\infty}^{\infty} \int \frac{dE dk_z dk_\perp k_\perp}{(2\pi)^3}
\phi(x) \phi^\dagger(x').
\label{GforS}
\end{eqnarray}
Therefore, substituting Eq.~(\ref{explphi}) into Eq.~(\ref{GforS}) and performing the integral over $\tau$, the expression for the fermion propagator can be written as
\begin{eqnarray}
S(x,x')&=&\sum_{\ell=-\infty}^{\infty} \int \frac{dEdk_z k_{\perp}dk_{\perp}}{(2\pi)^3}
\Phi(\rho,\rho')\nonumber\\
&\times&
\frac{e^{-i (E-(\ell+1/2)\Omega)(t-t')} e^{i k_z(z-z')} e^{i \ell(\varphi-\varphi')}}{E^2-k_z^2-m^2-k_{\perp}^2+i\epsilon}, \nonumber \\
\label{propprev}
\end{eqnarray}
where
\begin{eqnarray} 
&&\Phi(\rho,\rho')\equiv \text{diag} [(E-k_z+m+ik_{\perp})J_\ell(k_\perp \rho) J_\ell(k_\perp \rho'), \nonumber \\
&& (E+k_z+m-ik_{\perp})J_{\ell+1}(k_\perp \rho) J_{\ell+1}(k_\perp \rho')e^{i (\varphi-\varphi')}, \nonumber \\
&& (-E+k_z+m-ik_{\perp})J_\ell(k_\perp \rho) J_\ell(k_\perp \rho'), \nonumber \\
&& (-E-k_z+m+ik_{\perp})J_{\ell+1}(k_\perp \rho) J_{\ell+1}(k_\perp \rho')e^{i (\varphi-\varphi')}], \nonumber \\
\end{eqnarray}
and we have implemented the change of variable $E \rightarrow E+\Omega(\ell +1/2)$. Notice that the propagator turns out to be diagonal in Lorentz space. Also, translational invariance is only lost in the transverse direction, for otherwise the propagator depends on the coordinate differences $(t-t')$, $(z-z')$ and $(\varphi - \varphi')$.

The expression for the propagator can be further reduced. Let us focus on one of the elements, the component $S_{11}(x,x')$. We use the partial translation invariance to write
\begin{eqnarray}
S_{11}(\rho,\rho',\varphi,z,t)&=&\sum_{\ell=-\infty}^{\infty} \int \frac{dEdk_z k_{\perp}dk_{\perp}}{(2\pi)^3} \ \nonumber \\
&\times& (E-k_z+m+ik_{\perp})J_\ell(k_\perp \rho) J_\ell(k_\perp \rho')\nonumber \\
&\times& \frac{e^{-i (E-(\ell+1/2)\Omega)t}e^{i k_zz} e^{i \ell\varphi}}{E^2-k_z^2-m^2-k_{\perp}^2+i\epsilon}.
\end{eqnarray}
In order to calculate the sum over $\ell$, we use the integral representation of the Bessel functions
\begin{equation}
    J_\ell(x)=\frac{1}{2 \pi} \int_{-\pi}^\pi e^{i(x \sin(\tau)-\ell \tau)} d\tau,
\end{equation}
thus arriving at
\begin{eqnarray}
S_{11}(\rho,\rho',\varphi,z,t)&=&\int \frac{dEdk_z k_{\perp}dk_{\perp}}{(2\pi)^4}e^{-i(E-\Omega/2)t}e^{i k_zz}\nonumber\\
&\times&\frac{(E-k_z+m+ik_{\perp})}{E^2-k_z^2-m^2-k_{\perp}^2+i\epsilon}
\nonumber\\
&\times&\int_{-\pi}^\pi d\tau e^{i k_\perp \rho' \sin(\tau)}\nonumber\\
&\times&\sum_{\ell=-\infty}^\infty
J_\ell(k_\perp \rho) e^{i \ell(\varphi+\Omega t-\tau)}.
\label{propxxprime}
\end{eqnarray}
We now use the Jacobi-Anger expansion
\begin{equation}
    \sum_{\ell=-\infty}^{\infty}J_\ell(x)e^{i\ell y}=e^{i x \sin(y)},
\end{equation}
supplemented by the change of variable $\rho,\rho'\to R,r$ given by
\begin{eqnarray}
\rho'&=&R-r/2,\nonumber\\ 
\rho&=&R+r/2,
\label{changevar}
\end{eqnarray}
we get
\begin{eqnarray}
S_{11}(R,r,\varphi,z,t)&=&\int \frac{dEdk_z k_{\perp}dk_{\perp}}{(2\pi)^4}e^{-i(E-\Omega/2)t}e^{i k_zz}\nonumber\\
&\times&\frac{(E-k_z+m+ik_{\perp})}{E^2-k_z^2-m^2-k_{\perp}^2+i\epsilon}
\nonumber\\
&\times& \int_{-\pi}^\pi d\tau e^{-ik_\perp r(\sin\tau-\sin\theta)/2}\nonumber\\
&\times&
e^{ik_\perp R (\sin\tau+\sin\theta)},
\label{SafterrR}
\end{eqnarray}
where we have defined $\theta\equiv\varphi+\Omega t -\tau$.

We now make the approximation whereby the fermion is totally dragged by the vortical motion such that the angular position is  determined by the product of the angular velocity and the time, namely $\varphi+\Omega t=0$. This is a very good approximation, for instance, during the early stages of a peripheral heavy-ion collision, where particle interactions have not yet produced the development of a radial expansion, characterized by a rate $\Gamma$. In this sense, $\Omega$ being much larger than $\Gamma$, can be considered as the largest of the intrinsic energy scales in the problem. In this way, $\sin\theta \rightarrow -\sin\tau$ and thus the last factor in Eq.~(\ref{SafterrR}) becomes 1. Notice that, under this approximation, the function depends only on relative coordinates making it translationally invariant.

Using the identity
\begin{eqnarray}
\int_{-\pi}^\pi d\tau e^{ik_\perp r\sin\tau}=(2\pi)J_0(k_\perp r),
\label{J0}
\end{eqnarray}
we obtain
\begin{eqnarray}
\!\!\!\!\!S_{11}(r,\varphi,z,t)&=&\int \frac{dEdk_z k_{\perp}dk_{\perp}}{(2\pi)^3}e^{-i(E-\Omega/2)t}e^{i k_zz}\nonumber\\
&\times&\frac{(E-k_z+m+ik_{\perp})}{E^2-k_z^2-m^2-k_{\perp}^2+i\epsilon}J_0(k_\perp r).
\label{withJ0}
\end{eqnarray}
We now use that the function depends only on relative coordinates to introduce the Fourier transform
\begin{equation}
S_{11}(p)= \int d^4x e^{i p \cdot  x} S_{11}(x),
\end{equation}
and obtain
\begin{eqnarray}
S_{11}(p)=\frac{p_0+\Omega/2 -p_z+m+i p_{\perp}}{(p_0+\Omega/2)^2-\vec{p}^2-m^2+i\epsilon}.
\label{S11fin}
\end{eqnarray}
The rest of the terms in the propagator can be worked in a similar fashion and thus we get
\begin{widetext}
\begin{equation}
    S(p) = \begin{pmatrix}
\frac{p_0+\Omega/2 -p_z+m+i p_{\perp}}{(p_0+\Omega/2)^2-\vec{p}^2-m^2+i\epsilon} & 0  & 0 & 0 \\ 
0 & \frac{p_0-\Omega/2 +p_z+m-i p_{\perp}}{(p_0-\Omega/2)^2-\vec{p}^2-m^2+i\epsilon} & 0 & 0 \\
0 & 0 & \frac{-(p_0+\Omega/2) +p_z+m-i p_{\perp}}{(p_0+\Omega/2)^2-\vec{p}^2-m^2+i\epsilon} & 0 \\
0 & 0 & 0 & \frac{-(p_0-\Omega/2) -p_z+m+i p_{\perp}}{(p_0-\Omega/2)^2-\vec{p}^2-m^2+i\epsilon}\end{pmatrix}.
\end{equation}
\end{widetext}
The result can be further simplified by introducing the operators
\begin{equation}
\mathcal{O}^{\pm}\equiv\frac{1}{2}\left[1\pm i\gamma^1\gamma^2\right],
\end{equation}
such that the propagator looks like
\begin{eqnarray}
S(p)&=& \frac{[p_0+\Omega/2 -p_z+i p_{\perp}]\gamma_0+m}{(p_0+\Omega/2)^2-\vec{p}^2-m^2+i\epsilon} \mathcal{O}^{+} \nonumber \\
&+& \frac{[p_0-\Omega/2 +p_z-i p_{\perp}]\gamma_0+m}{(p_0-\Omega/2)^2-\vec{p}^2-m^2+i\epsilon}  \mathcal{O}^{-}\;.
\label{propwithO}
\end{eqnarray}
Equation~(\ref{propwithO}) is our main result. We emphasize that this propagator is obtained under the approximation whereby the fermion is dragged by the vortical motion. In this manner we have traded the detailed description of the fermion motion in favor of accounting for the way the angular velocity translates into an influence on the fermion spin degrees of freedom.

\section{Summary and outlook}\label{Concl}

In this work we have derived the propagator for a fermion immersed within a rigidly rotating environment. The motivation stems from the search of the description from first principles to study how the fermion spin is affected by the overall rotational motion. The method we used has been recently put forward in Refs.~\cite{Iablokov:2019rpd,Iablokov:2020upc} and it has been applied to rederiving the propagator for electrically charged bosons, fermions and even gauge bosons in the presence of a magnetic field. To our knowledge, this is the first time the method is used in the context of fermions immersed in a rotating environment. 

We found that the propagator is diagonal in Lorentz space and the general expression is not translationally invariant in the transverse radial coordinate. However, under the approximation that the fermion is completely dragged by the overall vortical motion,  translation invariance is recovered, which allows us to find the expression for the propagator in momentum space. 

The propagator thus found is now suited to be used in perturbative calculations using ordinary Feynman rules in momentum space. Work in this direction is currently being pursued and will be soon reported elsewhere.

\section*{Acknowledgments}

This work was supported in part by UNAM-DGAPA-PAPIIT Grant No. IG100219 and by Consejo Nacional de Ciencia y Tecnolog\'{\i}a Grants No. A1-S-7655 and No. A1-S-16215. R. Z. acknowledges support from   ANID/CONICYT FONDECYT Regular (Chile) under Grant No. 1200483.

%
%


\bibliography{bibliography}

\end{document}